\documentstyle[12pt,epsf,epsfig]{article}
\def\be{\begin{equation}}
\def\ee{\end{equation}}
\def\ba{\begin{eqnarray}}
\def\ea{\end{eqnarray}}

\def\bdm{\begin{displaymath}}
\def\edm{\end{displaymath}}
\def\la{~\mbox{\raisebox{-.6ex}{$\stackrel{<}{\sim}$}}~}
\def\ga{~\mbox{\raisebox{-.6ex}{$\stackrel{>}{\sim}$}}~}
\def\bq{\begin{quote}}
\def\eq{\end{quote}}

 at 10truept

\newcommand{\beq}{\begin{equation}}
\newcommand{\eeq}{\end{equation}}
\newcommand{\beqa}{\begin{eqnarray}}
\newcommand{\eeqa}{\end{eqnarray}}

\def\la{~\mbox{\raisebox{-.6ex}{$\stackrel{<}{\sim}$}}~}
\def\ga{~\mbox{\raisebox{-.6ex}{$\stackrel{>}{\sim}$}}~}

\def\ltap{\ \raise.3ex\hbox{$<$\kern-.75em\lower1ex\hbox{$\sim$}}\ }
\def\gtap{\ \raise.3ex\hbox{$>$\kern-.75em\lower1ex\hbox{$\sim$}}\ }
\def\gl{\ \raise.5ex\hbox{$>$}\kern-.8em\lower.5ex\hbox{$<$}\ }
\def\roughly#1{\raise.3ex\hbox{$#1$\kern-.75em\lower1ex\hbox{$\sim$}}}

\begin{document}

\thispagestyle{empty}
\begin{flushright}
hep-th/0702206\\ February 2007
\end{flushright}
\vspace*{1.2cm}
\begin{center}
{\Large \bf A New Dimension Hidden in the Shadow of a Wall}\\

\vspace*{2cm} {\large Nemanja Kaloper\footnote{\tt
kaloper@physics.ucdavis.edu}}\\
\vspace{.5cm} {\em Department of Physics, University of
California, Davis,
CA 95616}\\
\vspace{.15cm} \vspace{1.5cm} ABSTRACT
\end{center}

We propose a new way to hide the fifth dimension, and to modify
gravity in the far infra-red. A gravitating tensional membrane in
five dimensions folds the transverse space into a truncated cone,
stoppered by the membrane. For near-critical tension, the conical
opening is tiny, and the space becomes a very narrow conical
sliver. A very long section, of length comparable to the
membrane radius divided by the remaining conical angle, of this sliver 
is well approximated by a narrow cylinder ending on the membrane. Inside
this cylindrical throat we can reduce the theory on the circle. At
distances between the circle radius and the length of the
cylinder, the theory looks $4D$, with a Brans-Dicke-like gravity, and a preferred
direction, while at larger distances the cone opens up and the theory turns
$5D$. The gravitational light scalar in the throat can get an
effective local mass term from the interplay of matter
interactions and quantum effective potentials on the cone, which
may suppress its long range effects. We discuss some
phenomenologically interesting consequences.

\vfill \setcounter{page}{0} \setcounter{footnote}{0}
\newpage

In this note, we outline a novel mechanism to hide an extra
dimension and to change gravity at very large distances. It draws
on the fact that gravitating tensional codimension-2 objects fold
two transverse dimensions into a conical space. We will focus on
membranes in $5D$, but other examples abound, such as point masses
in $3D$ \cite{deser}, local cosmic strings in $4D$ \cite{strings},
or $3$-branes in $6D$ \cite{raman}. In realistic cases when
membranes are thick, their finite core resolves the tip of the
cone, which truncates on the membrane. For near-critical tensions
$\lambda \rightarrow 2\pi M_5^3$, the opening of the cone is tiny,
and the space looks like a semi-infinite conical sliver. Tuning
this may be easier than tuning the tension to zero. The
cosmological constant problem may in fact {\it help}. In field
theory, quantum corrections drive vacuum energy up to the UV
cutoff \cite{wein}. A field theory in the membrane core, and on
it, will generate large corrections to the tension. If its UV
cutoff is close to $5D$ Planck scale, the tension may be close
to critical. Alternatively, membranes could come with a full range
of tensions, some infinitesimally close to the critical value.
Here we will explore the consequences of near-critical tension.

The conical sliver is well approximated by a thin cylinder, of a
radius $r_0$ set by the membrane thickness, out to distances from
the membrane of the order of $r_0$ divided by the remaining
conical opening. This scale is our crossover scale. If we live on
this space, away from the membrane but inside the cylinder at
distances between $r_0$ and the crossover scale we won't see the
fifth dimension. The membrane is a domain wall, sitting at an end
of the world, its tension `propping up' the compact dimension.
Inside the cylindrical throat ending on the membrane, the theory
looks $4D$, with towers of heavy KK modes and a light
gravitational sector which contains the usual General Relativity,
a Brans-Dicke-like scalar, and a KK vector. The heavy KK states
are separated from the light modes by a mass gap set by the
membrane radius, $m_{\tt g} \sim 1/r_0$, and so are strongly
Yukawa-suppressed inside the cylinder. The KK vectors don't couple
to light matter directly. The scalar has a non-vanishing {\it vev}
$\Phi \sim r_0$ and varies very slowly along the direction normal
to the wall, memorizing that the background is really a $5D$ cone.
Its {\it vev} sets $4D$ Planck mass, by the usual Gauss law
$M_4^2 = M_5^3\Phi \sim M_5^3 r_0$. In vacuum, its fluctuations
would couple gravitationally to matter. However, inside matter
distributions this field may be screened by a combination of the
environmental effects \cite{dapol,khuwe} and quantum-mechanical
effective potentials \cite{conepots}.

Farther out the cone opens up. In dimensionally reduced theory,
this shows up as the scalar field increases with distance, making
$4D$ gravity weaker and the KK mass gap smaller. We no longer can
ignore the gravitons moving around the circle, and their
contributions begin to change the force. Eventually these infra-red
modifications change the theory back to a $5D$ theory on a
cone. This could have interesting observational consequences.

We start out with a thin membrane in an empty $5D$ spacetime,
with a tension $\lambda$. Its gravitational equations
are, in a membrane-fixed Gaussian-normal gauge
\cite{raman,kalkil},
\be M_5^3 \, G_5{}^A{}_B = - \lambda \, \delta^\alpha{}_\beta \,
\delta^A{}_\alpha \, \delta^\beta{}_B \, \delta^{(2)}(\vec{y}) \,
. \label{fieldeqs} \ee
The indices $A, B, \ldots$ run over $5D$ and $\alpha, \beta,
\ldots$ run over the $3D$ membrane worldvolume, $G_{AB} $ is the
$5D$ Einstein tensor, $\delta$-function is the tensor
$\delta^{(2)}(\vec{y}) = \frac{\sqrt{g_3}}{\sqrt{g_5}} \Pi
\delta(y_i)$,  and $(y_1,y_2)$ coordinatize the transverse space.
A flat membrane, with metric $g_{\alpha\beta} =
\eta_{\alpha\beta}$, is a solution of Eqs. (\ref{fieldeqs}) when
the transverse space is a cone \cite{raman}, whose metric is
\be ds_5^2 = \eta_{\alpha\beta} \, dx^\alpha dx^\beta +  dr^2 +
(1- \frac{\lambda}{2\pi M_5^{3}} )^2 r^2 d\phi^2 \, .
\label{5dcone} \ee
The deficit angle depends on the tension as $\Delta \phi =
\frac{\lambda}{M_5^{3}}$. The solution (\ref{5dcone}) is not well
defined at the critical value of the tension $\lambda_{cr} = 2\pi
M_5^3$, when it becomes degenerate. It is not clear what
(\ref{5dcone}) describes for the super-critical tensions $\lambda>
2\pi M_5^3$, since there is no limit connecting such solutions to
the vacuum. This puzzle also appears for cosmic strings in $4D$.
It is resolved by regulating the thin string with a finite core
\cite{naked}.

We do the same here, and replace the thin membrane by a tensional
3-brane wrapped on a circle, as in braneworld setups
\cite{msled,pesota,kalkil2}. To wrap a tensional 3-brane up and
make it look like a hollow membrane, we must cancel the pressure
$\propto \lambda_4$ on the circle. For this we put an axion
field $\Sigma$ on the 3-brane, whose vacuum action is
\be S_{vacuum} = - \int d^4 x \sqrt{g_4}\bigg(\lambda_4 +
\frac{1}{2} (\partial \Sigma)^2 \bigg) \, . \label{3brvac} \ee
Since the vacuum stress energy tensor is $T^a{}_b = - \lambda_5
\delta^a{}_b + \partial^a\Sigma \partial_b \Sigma -\frac{1}{2}
\delta^a{}_b g^{cd}\partial_c \Sigma \partial_d \Sigma$, with
lower case latin indices running over the 3-brane worldvolume, we
can take $\Sigma_0 = q \phi$ and pick the `charge' $q$ to obey
$q^2 = 2 r_0^2 \lambda_4$, where ${r_0}$ is the radius of the
cylindrical brane. This cancels the tensional pressure on the
circle. In a more realistic model of a thick membrane, this tuning
of $q$ v.s. $r_0$ should be replaced by a calculation of $r_0$ for
a given field configuration that resolves the core. For a thin
membrane we rewrite $\delta^{(2)}(\vec y)$ in polar
coordinates, using axial symmetry, as $\delta^{(2)}(\vec y)  =
\frac{1}{2\pi r} \delta(r)$. Shifting the argument to $r-r_0$ to
model a thick hollow membrane, with the source $\delta^{(2)}_{\rm
thick}(\vec y) = \frac{1}{2\pi r_0} \delta(r - r_0)$ we are led to
identify $\lambda = 4\pi r_0 \lambda_4$. The effective membrane
tension $\lambda$ includes the contributions of the axion
$\Sigma$. The field equations (\ref{fieldeqs}) thus change to
\be M_5^3 \, G_5{}^A{}_B  =  T^a{}_b \, \delta^A{}_a \,
\delta^b{}_B \,  \delta(r-r_0) \, .
\label{thickfieldeqs}
\ee
The stress energy tensor $T^a{}_b = - \frac{\lambda}{2\pi r_0}
\delta(r-r_0) \, {\rm diag}(1, 1, 1, 0)$ is covariantly conserved,
and supports a solution which looks like a truncated cone for
small tensions, ending on the brane at $r = r_0$ \cite{kalkil2}.
If we introduce a parameter $\varepsilon$ which measures the
deficit angle according to $\varepsilon = 1-\frac{\lambda}{2\pi
M_5^3}$, such that $\Delta \phi = 2\pi (1-\varepsilon)$, we can
write the metric solution as
\be ds_5{}^2 = \eta_{\alpha\beta} \, dx^\alpha dx^\beta + dr^2
+\bigg[ \Bigl(1-(1-\varepsilon) \Theta(r-r_0)\Bigr)r +
(1-\varepsilon) r_0 \Theta(r-r_0)\bigg]^2 d\phi^2 \, ,
\label{thick5dvac} \ee
where $\Theta(x)$ is the Heaviside step function.

In the critical membrane limit $\lambda_c = 2\pi M_5^3$,  so that
$\varepsilon = 0$. In this limit, the exterior geometry of
(\ref{thick5dvac}) deforms into a semi-infinite cylinder of
constant radius, equal to the membrane thickness. This cylindrical
throat looks like a $4D$ spacetime, at all distances $\ell > r_0$,
compactified by the critical membrane tension. On the other hand,
the supercritical solutions $\varepsilon < 0$ are singular at $r =
\frac{1+|\varepsilon|}{|\varepsilon|} r_0 > r_0$, outside of the
membrane, and so they spontaneously compactify on a $2D$ teardrop,
similar to \cite{gell}. So supercritical vacua (\ref{thick5dvac})
look like $3D$ spacetimes. Therefore static, flat, lonely branes
in infinite space only occur for $\lambda \le \lambda_c$.

\begin{figure}[thb]
\vskip.3cm
\centerline{\includegraphics[width=0.4\hsize,width=0.3
\vsize,angle=0]{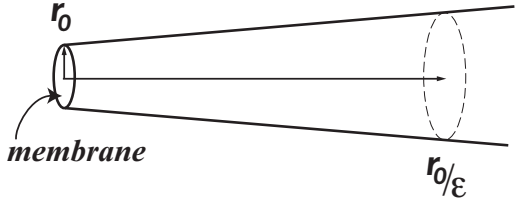}}
\caption{{$2D$ conical transverse geometry of a thick membrane vacuum.}
\label{figure}}
\end{figure}
Here we will be most interested in near-critical membranes, with
$0<\varepsilon \ll 1$. Their surroundings looks like a truncated conical
sliver in Fig. (\ref{figure}). The exterior metric is
\be ds_5{}^2 = \eta_{\alpha\beta} \, dx^\alpha dx^\beta + dr^2 +
\bigg(r_0 + \varepsilon (r - r_0)\bigg)^2\ d\phi^2 \, .
\label{sliver} \ee
It approximates a cylinder for distances $r_0 \la \ell \la
r_0/\varepsilon$, because the radius of the sliver changes very
little in this regime. To make the cylindrical throat in Fig.
(\ref{figure}) very long, the tension must be very close to the
critical tension $\lambda_{cr} = 2\pi M_5^3$, to get $\varepsilon
\ll 1$.

Since the vacuum (\ref{thick5dvac}) is axially symmetric, we can
dimensionally reduce the $5D$ theory to $4D$, and explore it using
effective field theory methods. For simplicity, we take
\be S = \int d^5x \sqrt{g_5} \bigg( \frac{M_5^3}{2} R_5 - {\cal
L}_{\rm matter} \bigg) - \int_{r = r_0} d^4 x \sqrt{g_4}
\bigg(\lambda_4 + \frac12 (\partial \Sigma)^2 \bigg) + {\rm
boundary ~ terms} \, , \label{5daction} \ee
which describes full $5D$ gravity and matter, and a wrapped
3-brane of radius $r_0$, with tension and axion $\Sigma$. We could
also add another 3-brane, orthogonal to the
tensional brane in (\ref{5daction}), to localize additional
light $4D$ matter fields, and avoid matter KK partners.
For the time being we will work with (\ref{5daction}), but will
elaborate this later on. In (\ref{5daction}), the boundary terms
covariantize membrane's gravitational couplings. Any axially
symmetric deformations of (\ref{thick5dvac}), that describe
axially symmetric fluctuations of the radius of the sliver and its
shear relative to the membrane, will play the role of light fields
in the gravitational sector upon reduction to $4D$. They are
encoded in
\be ds_5{}^2 = g_{\mu\nu}(x^\mu) \, dx^\mu dx^\nu + \Phi^2(x^\mu)
\Bigl(d\phi - V_\mu(x^\mu) dx^\mu\Bigr){}^2  \, .
\label{pert5dvac} \ee
Here we have {\it recombined} the longitudinal membrane
coordinates $x^\alpha$ and the transverse coordinate $r$ together,
into a chart of a $4D$ spacetime: $x^\mu = \{x^\alpha, r\}$.
Treating $\phi$-dependence of all the fields perturbatively,
and expanding them in Fourier series
\be \Psi_{\{{\cal N}\}} (x^\mu,  r_0  \phi) = \sum^\infty_{n = -
\infty} \Psi_{\{{\cal N}\}, n}(x^\mu) \, e^{i n \phi} \, ,
\label{KK} \ee
we can rewrite (\ref{5daction}) as a $4D$ theory with KK towers of
fields,  with masses $M^2 = m^2 + n^2 m_{\tt g}^2$, where $m$ are
$5D$ mass terms and $m_{\tt g}$ are mass gaps. The label
$\{{\cal N}\}$ denotes different $4D$ representations of $5D$
fields. For membrane-borne fields, the mass gap is just $m_{\tt g}
= 1/r_0$. For the $5D$ matter fields, the mass gap is set by the
radius of the circle to $m_{\tt g} =\Phi_0{}^{-1}$. Inside the
cylindrical throat, at distances $r_0 \la \ell \la
r_0/\varepsilon$, the gap is $m_{\tt g} \sim 1/r_0$ and these
states will be strongly Yukawa-suppressed, $\propto e^{-M\ell}
\ll 1$. So in this regime we can neglect all the heavy KK states.
The light ones behave as our $4D$ degrees of freedom. As
distance increases, the background value $\Phi_0$ grows, and so the gap
eventually disappears. At very large distances $\ell \gg r_0/\varepsilon$
we can't ignore the KK towers any more. In that
limit, the theory reveals the fifth dimension.

Thus inside the cylindrical throat (\ref{thick5dvac}) we can
truncate (\ref{5daction}) to only the light fields, and get
\ba
S_{4D \, {\rm eff}} &=&  \int d^4x \sqrt{g_4}
\bigg( \pi M_5^3 \Phi \, R_4 -
\frac14 \Phi F_{\mu\nu}^2 -
2\pi \Phi \, {\cal L}_{\rm matter} \bigg) \nonumber \\
&-& \int_{r = r_0} d^3 x  \sqrt{g_3} \bigg(\frac{\lambda}{2}
\Bigl(\frac{\Phi}{r_0} +  \frac{r_0}{\Phi} \Bigr) + \pi {\Phi}
(D_\alpha \sigma)^2 \bigg) + {\rm boundary ~ terms} \, ,
\label{4daction} \ea
in the membrane-fixed gauge. Here $F_{\mu\nu}$ is the field
strength of the Abelian KK vector field $A_\mu = (\pi
M_5^3)^{1/2}V_\mu$, $\sigma = \Sigma - \Sigma_0$ is the lightest,
axially symmetric, axion field fluctuation on the membrane, and
$D_\alpha \sigma = \partial_\alpha \sigma +q A_\alpha/ (\pi
M_5^3)^{1/2}$ is its St\"uckelberg  gauge-covariant derivative. In
evaluating (\ref{4daction}) we have used $\Sigma_0 = q \phi$,
where $q^2 = 2 r_0^2 \lambda_4$, and the relationship between
$\lambda$ and $\lambda_4$ given in the text before Eq.
(\ref{thickfieldeqs}). The matter fields in ${\cal L}_{\rm
matter}$ couple minimally to the metric $g_{\mu\nu}$ and
multiplicatively to $\Phi$, as is manifest in (\ref{4daction}).
The light fields, except $\sigma$, are KK gauge singlets. The
heavy KK states are not, but in the regime we are exploring they
play no role. From  (\ref{4daction}), $\sigma$ is a St\" uckelberg
field of $A_\mu$, localized on the membrane. In a unitary gauge,
given by $\hat A_\mu =  A_\mu + (\pi M_5^3)^{1/2}
\partial_\mu \sigma/q$, this term is just the membrane-localized mass term for
$\hat A_\mu$. So light matter is decoupled from $\hat A_\mu$
in the leading order, and we can drop it from further consideration as mostly
harmless.

This leaves the graviton and the light Brans-Dicke-like scalar
$\Phi$, whose action reduces to
\be S_{4D \, {\rm eff}} = \int d^4x \sqrt{g_4} \bigg( \pi M_5^3
\Phi \, R_4 - 2\pi \Phi \, {\cal L}_{\rm matter} \bigg) - \,
\frac{\lambda}{2} \int_{r = r_0} d^3 x \sqrt{g_3}  \,
\Bigl(\frac{\Phi}{r_0} +  \frac{r_0}{\Phi} \Bigr) \, + \, {\rm
boundary ~ terms} \, . \label{4BD} \ee
This is a scalar-tensor gravity non-minimally coupled to matter
and to the membrane. At distances $\ell > r_0$ the membrane looks
like a domain wall at the end of the universe. When the matter
fields are in the vacuum, so ${\cal L}_{\rm matter} = 0$,  the
field equations which come from the action (\ref{4BD}) admit the
flat space solution $g_{0 \, \mu\nu} = \eta_{\mu\nu}$ when
$\Phi|_{r = r_0} = r_0$, reducing to a single equation for the
scalar field $\pi M_5^3 \partial^2 \Phi = - \frac12 \lambda
\delta(r - r_0)$. Choosing $\Phi|_{r=0} = 0$ by using the Gauss
law inside the membrane, we recover $\Phi_0 =
\Bigl(1-(1-\varepsilon) \Theta(r-r_0)\Bigr)r + (1-\varepsilon) r_0
\Theta(r-r_0)$, as expected. We focus only on the boundary
conditions inherited from $5D$ because (\ref{4BD}) is just a tool
to study the IR dynamics of the original $5D$ theory.

To see how matter gravitates in the throat, we go to the Einstein
frame, corresponding to the normal modes description of
(\ref{4BD}). We define the dimensionless scalar $\hat \Phi =
\Phi/r_0$ and $4D$ Planck mass $M_4^2 = 2\pi M_5^3 r_0$, and
go to the new variables by $\hat \Phi = \exp({\sqrt{\frac23
}\frac{\varphi}{M_4}} )$, and $g_{\mu\nu} = \exp(-{\sqrt{\frac23 }
\frac{\varphi}{M_4}}) \, \bar g_{\mu\nu}$. Redefining the matter
Lagrangian by $2\pi r_0 \, {\cal L}_{\rm matter} \rightarrow {\cal
L}_{\rm matter}$, we get
\ba S_{4D \, {\rm eff}} &=& \int d^4 x \sqrt{\bar g_4} \, \bigg(
\frac{M^2_4}{2} \bar R_4 - \frac12 (\bar \nabla \varphi)^2 -
e^{-\sqrt{\frac23}\frac{\varphi}{M_4}} {\cal L}_{\rm
matter}(e^{-\sqrt{\frac23}\frac{\varphi}{M_4}} \bar g_{\mu\nu})
\bigg) \nonumber \\
&-& \, \frac{\lambda}{2} \int_{r = r_0} d^3 x \sqrt{\bar g_3}  \,
e^{-\sqrt{\frac32 } \frac{\varphi}{M_4}} \,
\Bigl(e^{-\sqrt{\frac23 } \frac{\varphi}{M_4}}  + e^{\sqrt{\frac23
} \frac{\varphi}{M_4}}  \Bigr) \, + \, {\rm boundary ~ terms} \, .
\label{ebdacts} \ea
To get the background solutions for the Einstein frame fields
$\varphi$ and $\bar g_{\mu\nu}$ we substitute in the redefinitions
the expressions for $\hat \Phi_0$ and $g_{0 \, \mu\nu} =
\eta_{\mu\nu}$. The short-distance singularities in  these
variables, present because $\exp({\sqrt{\frac23 }
\frac{\varphi_0}{M_4}})  \sim r/r_0$ and $d\bar s^2_0 \sim
\frac{r}{r_0} (\, \eta_{\alpha\beta} dx^\alpha dx^\beta + dr^2)$,
are harmless, because the scalar is really a shrinking polar
radius of the fifth dimension, describing how the full $5D$ metric
approaches Minkowski space in the membrane core.

As it stands, the theory (\ref{ebdacts}) is phenomenologically
problematic. To see why, consider the scalar field equation. By
varying (\ref{ebdacts}), and denoting the membrane term by
$\lambda(\varphi)$, it is
\be \nabla^2 \varphi \, = \, \frac{\partial
\lambda(\varphi)}{\partial \varphi} \, \sqrt{\frac{{\bar
g_3}}{{\bar g_4}} } \delta(r-r_0)  \,  + \, \frac{1}{\sqrt{6} M_4}
e^{-\sqrt{\frac23 }\frac{\varphi}{M_4}} \Bigl(\bar T + 2{\cal
L}_{\rm matter} \Bigr) \, , \label{bdfieq} \ee
where $\bar T $ is the trace of the matter stress energy tensor,
defined by the variation of the matter action as  $\delta S_{\rm
matter} = \frac12 \int d^4x \sqrt{\bar g_4} \,
\exp({-\sqrt{\frac23 } \frac{\varphi}{M_4}}) \, \bar T^{\mu\nu} \,
\delta \bar g_{\mu\nu}$. The Lagrangian term ${\cal L}_{\rm
matter}$ appears due to non-minimal couplings in the action
(\ref{ebdacts}), as is known from \cite{wetterich}, but can be
handled easily. It behaves as pressure, and for non-relativistic
sources, it can be neglected relative to $\bar T$. The membrane
terms set field gradients in the vacuum, controlling the very long
range asymptotics far from matter thanks to the
codimension-1/codimension-2 membrane dynamics (after/before
reduction, respectively), and the fact that the fields outside of
local sources decrease with distance. On the other hand, they do
not play a significant role in the local source dynamics far from
the membrane. Neglecting them, we see that $\varphi$, as it stands
in (\ref{ebdacts}), is too light and gravitationally coupled.

However in physically realistic situations quantum corrections may
generate an effective potential for $\varphi$. For example, in
conical spaces quantum corrections may generate potentials, that
would depend on the logs of fields, including $V_{\rm eff} \sim
\ln(\Phi)^{-1}$ in $5D$ \cite{conepots}. After dimensional
reduction, and with the overall conical radius being held up by
the membrane, a relevant correction from such a potential may be
$V_{4D \, {\rm eff}} \sim - \mu^5 /\ln(\hat \Phi)$, where $\mu$ is
a scale which depends on the conical radius $r_0$ and $5D$
Planck scale $M_5$. In the Einstein frame, this may yield an extra
term $V^E_{4D \, {\rm eff}} = - \mu^{5} /(\varphi + v_0)$, for
some scale $v_0$, chosen so that this potential is small near the
classical vacuum $\varphi = 0$.

Further, we can use the conservation of the stress-energy tensor,
and apply it to non-relativistic sources. It is $\bar
\nabla_\mu(\exp({\sqrt{-\frac23 }\frac{\varphi}{M_4}})  \bar
T^{\mu\nu}) = - \frac{1}{\sqrt{6} M_4} \bar \nabla^\nu \varphi
\exp({\sqrt{-\frac23 }\frac{\varphi}{M_4}}) \bigl[\bar T + 2{\cal
L}_{\rm matter}\bigr]$. For non-relativistic sources, with our
conventions $\bar T^{00} \simeq \bar\rho$, while all other
components of stress energy tensor, and the Lagrangian itself,
vanish in the leading order. So the conservation equation yields
that the conserved quantity independent of $\varphi$, which in the
$\varphi$-equation plays the same role as the centrifugal barrier
term in the central force problem, is $\rho =
\exp(-\frac{\varphi}{\sqrt{6} M_4}) \bar \rho$. Then, after adding
the effective potential, dropping the membrane term and replacing
$\bar T^{\mu\nu}$ by $\propto \rho$ terms, Eq. (\ref{bdfieq}) near
mass distributions becomes
\be \nabla^2 \varphi \, =  \, - \frac{\rho}{\sqrt{6} M_4}
e^{-\frac{\varphi}{\sqrt{6}M_4}} + \frac{\partial V^E_{4D \, {\rm
eff}}}{\partial \varphi} \, . \label{bdfieqeff} \ee
Inside mass distributions, {\it vev} of $\varphi$ shifts from its
vacuum to a $\varphi_*$, where the right-hand side of
(\ref{bdfieqeff}) vanishes. Around this shifted background, small
perturbations of $\varphi$ become massive, with the effective mass
given by $m^2_{\rm eff} = \frac{\partial^2 V^E_{4D \, {\rm
eff}}}{\partial \varphi_*^2} \sim \frac{\rho^{3/2}}{M_4^{3/2}
\mu^{5/2}}$. This may screen most of the interior of the
distribution $\rho$ from enacting a large long-range scalar force,
allowing only a force from a thin outer layer of thickness
$1/m_{\rm eff}$ \cite{khuwe}. Also, the total shift of $\varphi$
from its vacuum value should be much smaller than $M_4$ in order
to avoid conflicts with bounds on the variation of Newton's
constant.

Having $\mu$ in this formula helps; it can't be too large since in
that case the screening mass for $\varphi$ would be too small. In
fact, to pass the terrestrial table-top bounds, for example, the
scale $\mu$ must obey, roughly, $\mu < 0.1 \, {\rm eV}$, implying
that the potential $V^E_{4D \, {\rm eff}}$ had better be very
small. While calculating this potential explicitly is beyond the
scope of the present work, there exist some ideas for how such
small potentials could be obtained
\cite{conepots,garriga,albrecht,pelpop}, after a suitable
subtraction of UV divergences. Other scenarios, where
gravitational light scalars decouple due to environmental effects,
were discussed in \cite{dapol}.

We get more bounds from requiring that the theory looks $4D$ at 
observationally accessible scales. If our matter
comes from the reduction of a $5D$ action, then KK
mass gap must be ${\cal O}({\rm TeV})$, or $r_0 \la 10^{-18} \, {\rm m}$, 
to conform with collider bounds. If we 
also require that gravity is $4D$ out
to the horizon scale (which may be too conservative), 
$r_0/\varepsilon \ga H_0^{-1} \sim 10^{26} \, {\rm m}$. These
bounds combine into $\varepsilon \la 10^{-44}$, implying that
the deficit angle must be extremely
close to $2\pi$. As we noted, this may occur if the UV cutoff of
the theory on the membrane is extremely close to $5D$ Planck
scale. If so, this value, once set, may not change so much by 
subsequent radiative corrections. In this case KK mass scale
is in the LHC reach, while $M_5 \sim 10^{14} \, {\rm GeV}$ is
close to $M_{\rm GUT}$. If we take an additional
3-brane in the bulk, with zero tension so it doesn't
disturb the background, and let the light matter including the Standard Model 
live on it, we can raise $r_0$ up to the
table-top gravity bound of $10^{-4} \, {\rm m}$. Then
$5D$ Planck scale is $M_5 \sim 10^{9} \, {\rm GeV}$. In this case,
matter couples differently to $\Phi$, but similar environmental
decoupling mechanisms may still neutralize the scalar force.

The background (\ref{thick5dvac}) has a preferred direction, given
by the outward normal to the membrane. As we move away from the
membrane, both the particle physics couplings, controlled by the
slowly varying $\Phi_0$, and the geometry will change. Thus
Lorentz and translational symmetries are broken by the membrane's
gravitational field. Similar Lorentz violations have been noted
recently in \cite{oriol}. In our example, this is how the
background memorizes that it is a part of a conical space in $5D$.
The breaking is tied to the scale of gravity modification, and so
it is small. However this opens up an interesting possibility,
that both Lorentz symmetry is broken and gravity is modified at
cosmological scales. One might use such backgrounds to model
cosmologies with a preferred direction, as in \cite{magoo}.
Alternatively, one may be able to constrain such Lorentz breaking
and gravity modification from cosmological data.

To sum up, we have outlined a novel way to hide a spatial
dimension at short and intermediate scales, but let it reopen at
cosmological scales. The theory at distances $r_0 \la \ell \la
r_0/\varepsilon$ looks like a scalar-tensor gravity with a
membrane at the end of the world. The scalar long range force may
be suppressed by a combination of environmental effects and an
effective potential from quantum corrections. In that case, the
theory may yield interesting correlations between gravity
modifications and Lorentz breaking at cosmological scales, while
remaining consistent with current bounds. It would be of interest
to explore this mechanism further.

%%%%%%%%%%%%%%%%%%%%%%%%%%%%%%%%%%%%%%%%%%%%%%%%%%
\vskip.5cm
%%%%%%%%%%%%%%%%%%%%%%%%%%%%%%%%%%%%%%%%%%%%%%%%%%%

{\bf \noindent Acknowledgements}

\smallskip

We thank S. Carlip, A. Iglesias, D. Kiley, M. Park, L. Sorbo and
J. Terning for useful discussions. This work was supported in part
by the DOE Grant DE-FG03-91ER40674, in part by the NSF Grant
PHY-0332258 and in part by a Research Innovation Award from the
Research Corporation.

%\pagebreak

\end{document}